\documentclass[12pt]{iopart}

\usepackage{graphicx}
\begin{document}

\title[Low-temperature giant coercivity in Co$_{6.2}$Ga$_{3.8-x}$Ge$_{x}$ ($x$=2.4 to 3.2)]{Low-temperature giant coercivity in Co$_{6.2}$Ga$_{3.8-x}$Ge$_{x}$ ($x$=2.4 to 3.2)}

\author{Jiro Kitagawa$^1$, Himawari Nomura$^1$ and Terukazu Nishizaki$^2$}

\address{$^1$ Department of Electrical Engineering, Faculty of Engineering, Fukuoka Institute of Technology, 3-30-1 Wajiro-higashi, Higashi-ku, Fukuoka, 811-0295, Japan}
\address{$^2$ Department of Electrical Engineering, Faculty of Science and Engineering, Kyushu Sangyo University, 2-3-1 Matsukadai, Higashi-ku, Fukuoka, 813-8503, Japan}
\ead{j-kitagawa@fit.ac.jp}

\begin{abstract}
The observation of giant coercivity exceeding 20 kOe at low temperatures in several transition-metal-based compounds has attracted significant attention from a fundamental perspective. This research is also relevant to developing rare-earth-free permanent magnets, wherein cobalt is one of the primary elements used. To facilitate easy fabrication, rare-earth-free and Co-based inorganic bulk magnets that exhibit giant coercivity are highly demanded but rarely reported. Herein, we report the observation of low-temperature giant coercivity in polycrystalline metallic Co$_{6.2}$Ga$_{3.8-x}$Ge$_{x}$ ($x$=2.4 to 3.2) with the hexagonal Fe$_{13}$Ge$_{8}$-type structure composed of Kagome and triangular lattices. As the Ge content $x$ decreases from 3.2, the magnetic ground state changes from ferrimagnetism to ferromagnetism at $x$=2.6. In the ferrimagnetic state, we observed a signature of spin frustration arising from the Kagome and/or triangular lattices of Co atoms. The ferromagnetic ordering temperatures for the $x$=2.6 and 2.4 samples are 46 K and 60 K, respectively. The coercive fields rapidly increase upon cooling and reach values of 26 kOe and 44 kOe in the $x$=2.6 and 2.4 samples, respectively, at 2 K. 
\end{abstract}

\noindent{\it keywords}: giant coercivity, geometrical frustration, magnetization

%
%
%
%
%

\section{Introduction}
NdFeB and Sm-Co permanent magnets are commonly used in modern electric society. 
However, due to the high supply risk associated with rare earth elements such as Nd and Sm, developing rare-earth-free permanent magnets has become a major research focus\cite{Cui:AM2018}. 
To be commercially viable, permanent magnets require a large coercive field $H_\mathrm{c}$ and high saturation magnetization. 
However, commercial rare-earth-free magnets such as Alnico and ferrite exhibit low $H_\mathrm{c}$ values of approximately 3 kOe, much lower than NdFeB and Sm-Co magnets with $H_\mathrm{c}$ values of 15-20 kOe\cite{Cui:AM2018}. 
Recently, some transition-metal-based magnets have been reported to exhibit huge coercivity values that exceed 20 kOe, known as giant coercivity\cite{Gorbachev:RCR2021}. 
Typically, giant coercivity is observed at low temperatures but often surpasses the $H_\mathrm{c}$ values of NdFeB and Sm-Co magnets. 
Rare-earth-free transition-metal oxides like Mn$_{2}$LiReO$_{6}$ and Sr$_{5}$Ru$_{4.1}$O$_{15}$ or some Fe-based compounds have been reported as giant coercive compounds\cite{Solana:JMC2022,Ymamamoto:CM2010,Zhang:CM2011,Negishi:JMMM1987,Debnath:JAP:2020,Bhattacharya:PRB2016,Majumder:JMMM2020}. 
In most cases, $H_\mathrm{c}$ increases rapidly on cooling and reaches values of 40-120 kOe at 2-4 K. 
The standard features among these giant coercive compounds are an anisotropic crystal structure, canted ferromagnetism, and low saturation magnetization. 
Despite the relatively high saturation magnetization exhibited by Alnico and ferrite magnets, understanding the underlying mechanism of giant coercivity would prove advantageous as it can inform the exploration of new rare-earth-free hard magnets.

One of the fundamental constituents of commercially-viable rare-earth-free magnets is cobalt. 
For facile fabrication, a bulk sample is highly desirable; however, reports of rare-earth-free and Co-based non-molecular inorganic bulk magnets exhibiting giant coercivity values over 20 kOe are few and far between. 
While it is true that room-temperature hard magnetic CoPt thin films, cobalt monolayers, and Co$_{3}$Sn$_{2}$S$_{2}$ wires have been shown to exhibit giant coercivity values\cite{Rakshit:APL2006,Lin:APL2012,Shiogai:PRM2022}, the only known example of bulk magnets would be CaBaCo$_{4}$O$_{7}$ with the Curie temperature $T_\mathrm{C}$ of 70 K and $H_\mathrm{c}$ of 20 kOe at 5 K, and K$_{2}$Co$_{3}$(OH)$_{2}$(SO$_{4}$)$_{3}$(H$_{2}$O)$_{2}$ with $T_\mathrm{C}$ = 30 K and $H_\mathrm{c}$ = 50 kOe at 1.8 K\cite{Caignaert:SSC2009,Vilminot:CM2010}. 
Thus, exploring new Co-based non-molecular inorganic bulk ferromagnets with giant coercivity is meaningful. 

\begin{figure}
\centering
\includegraphics[width=1.0\linewidth]{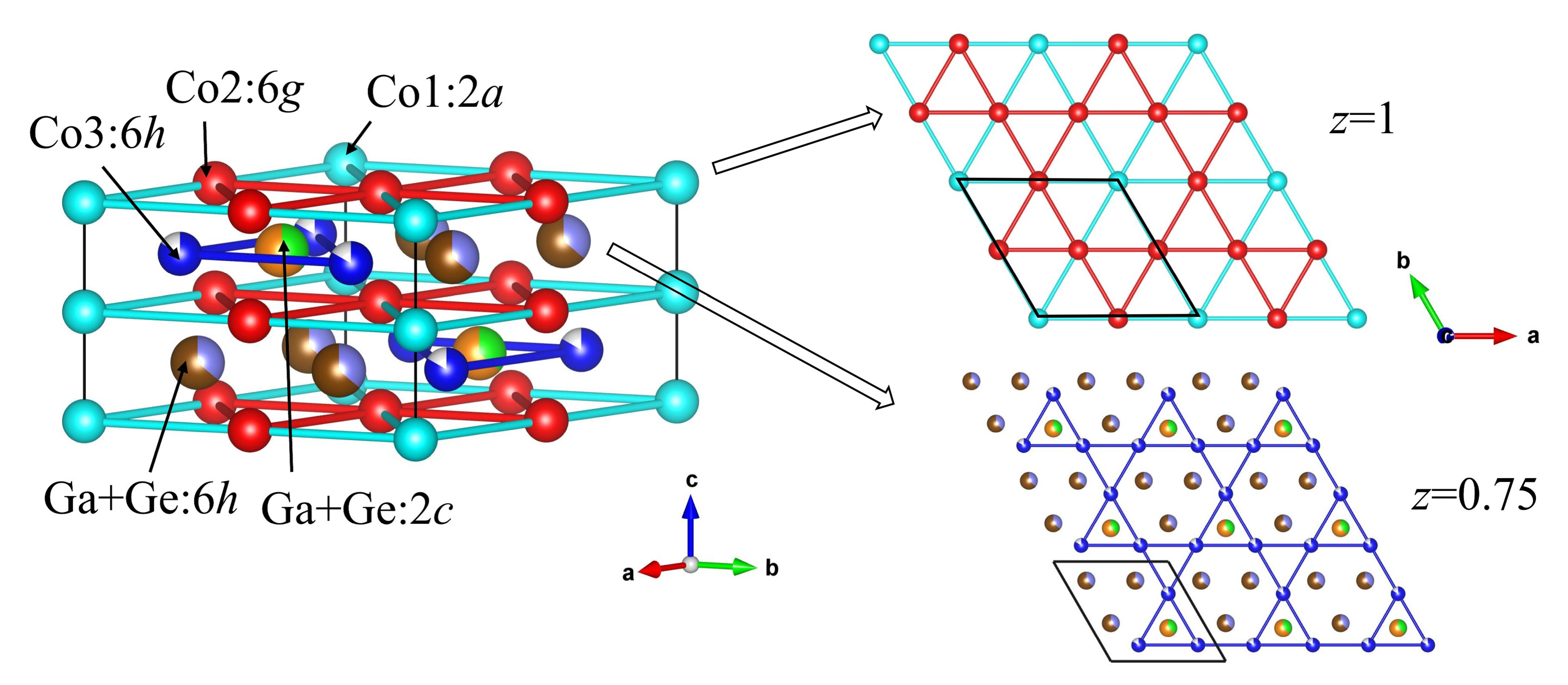}
\caption{\label{fig1} Crystal structure of Co$_{6.2}$Ga$_{3.8-x}$Ge$_{x}$. The solid black lines represent unit cell.}
\end{figure}

Our focus is on transition-metal-based compounds with the hexagonal Fe$_{13}$Ge$_{8}$-type structure, which have not been extensively investigated. 
Only the magnetic properties of Fe$_{3}$Ga$_{2-x}$As$_{x}$ (0.21 $\leq$ $x$ $\leq$ 0.85) and Fe$_{3}$Ga$_{0.35}$Ge$_{1.65}$ have been reported\cite{Moze:JPCM1994,Kitagawa:JPSJ2022}, and they exhibit soft ferromagnetism at room temperature. 
While Co$_{6.2}$Ga$_{3.8-x}$Ge$_{x}$ also crystallizes into the Fe$_{13}$Ge$_{8}$-type structure\cite{Malaman:JLCM1980}, its magnetic properties have not yet been studied. 
Figure \ref{fig1} illustrates the crystal structure, and Table \ref{tab1} provides details of the atomic positions. 
The space group is $P$6$_{3}$/$mmc$ (No. 194), and there are three Wyckoff sites 2$a$, 6$g$, and 6$h$ for Co atoms. 
The occupancy at the 6$h$ site is less than 1.0, which suggests the presence of vacancies. 
The Co2 and Co3 atoms form Kagome networks, while the Co1 atoms form a triangular lattice with a relatively long interatomic distance. 
We observe that the Co2 atoms form the ideal Kagome lattice, and the Co3 atoms form a slightly distorted Kagome lattice with vacancies. 
Ga and Ge atoms randomly occupy the 2$c$ and another 6$h$ sites.

In this article, we have examined the magnetic and transport properties of polycrystalline Co$_{6.2}$Ga$_{3.8-x}$Ge$_{x}$ with $x$ ranging from 2.4 to 3.2. 
Our investigation has revealed that the material exhibits a metallic nature with itinerant $d$-electrons of Co and displays a change from the ferromagnetic (FM) to the ferrimagnetic ground state with an increase in $x$. 
This crossover is accompanied by a shift towards a spin frustration regime dominated by antiferromagnetic (AFM) interaction. 
Remarkably, we have found that the giant coercivity values of 26 kOe and 44 kOe were achieved in Co$_{6.2}$Ga$_{1.2}$Ge$_{2.6}$ and Co$_{6.2}$Ga$_{1.4}$Ge$_{2.4}$, respectively, at a temperature of 2 K. Our study has further shown that the geometrical frustration with the itinerant $d$-electrons and canted spin structure, together with the anisotropic crystal structure, contribute to the emergence of the giant coercivity.

\begin{table}
\centering
\caption{\label{tab1}Atomic positions of Co$_{6.2}$Ga$_{3.8-x}$Ge$_{x}$.}
\begin{tabular}{cccccc}
\br
 Atom & Wyckoff & $x$ & $y$ & $z$ & Occupancy \\
\mr
Co1 & 2$a$ & 0 & 0 & 0 & 1   \\
Co2 & 6$g$ & 0.5 & 0 & 0 & 1   \\
Co3 & 6$h$ & 0.1614 & 0.3228 & 1/4 & 0.83   \\
Ga+Ge & 2$c$ & 1/3 & 2/3 & 1/4 & 1   \\
Ga+Ge & 6$h$ & 0.8088 & 0.6176 & 1/4 & 1   \\
\br
\end{tabular}
\end{table}

\section{Materials and Methods}
Polycrystalline specimens were synthesized via a homemade arc furnace employing constituent elements Co (99.99 \%), Ga (99.99 \%), and Ge (99.999 \%). 
The elements were co-melted through an arc-melting process to yield button-shaped samples weighing 1.5 g on a water-cooled Cu hearth. 
To ensure homogeneity, each sample was subjected to multiple flips and remelting. 
Subsequently, each as-cast sample was sealed in an evacuated quartz tube and annealed at 700 $^{\circ}$C for 4 days.

The X-ray diffraction (XRD) patterns of the powdered samples were collected at room temperature using a Bragg-Brentano geometry X-ray diffractometer (XRD-7000L, Shimadzu, Kyoto, Japan) with Cu-K$\alpha$ radiation. 
A field-emission scanning electron microscope (FE-SEM; JSM-7100F, JEOL, Akishima, Japan) was employed for the metallographic examination, and an energy-dispersive X-ray (EDX) spectrometer, which was equipped with the FE-SEM, was used for the atomic composition analysis.

The temperature dependence of the dc magnetic susceptibility, $\chi_\mathrm{dc}$ ($T$), ranging from 2 K to 300 K, and the isothermal magnetization curve were measured using MPMS3 (Quantum Design, San Diego, CA, USA). 
In the magnetization curve measurements, except for the samples with $x$= 2.4, 2.6, and 2.8 at 2 K, we employed demagnetization using the field oscillation mode in MPMS3 prior to each measurement. 
Despite this demagnetization procedure, a residual magnetization was observed when a high coercive field was obtained. 
Consequently, we excluded the initial magnetization process from the data display. 
To obtain precise initial magnetization curves, magnetization curves at 2 K were measured for samples with $x$ values of 2.4, 2.6, and 2.8 after subjecting them to zero-field cooling. 
The temperature dependence of electrical resistivity, $\rho$ ($T$), ranging from 3 K to room temperature, was measured using a dc four-probe method with a homemade system in a GM refrigerator (UW404, Ulvac cryogenics, Kyoto, Japan).

\section{Results and Discussion}
\begin{figure}
\centering
\includegraphics[width=1.0\linewidth]{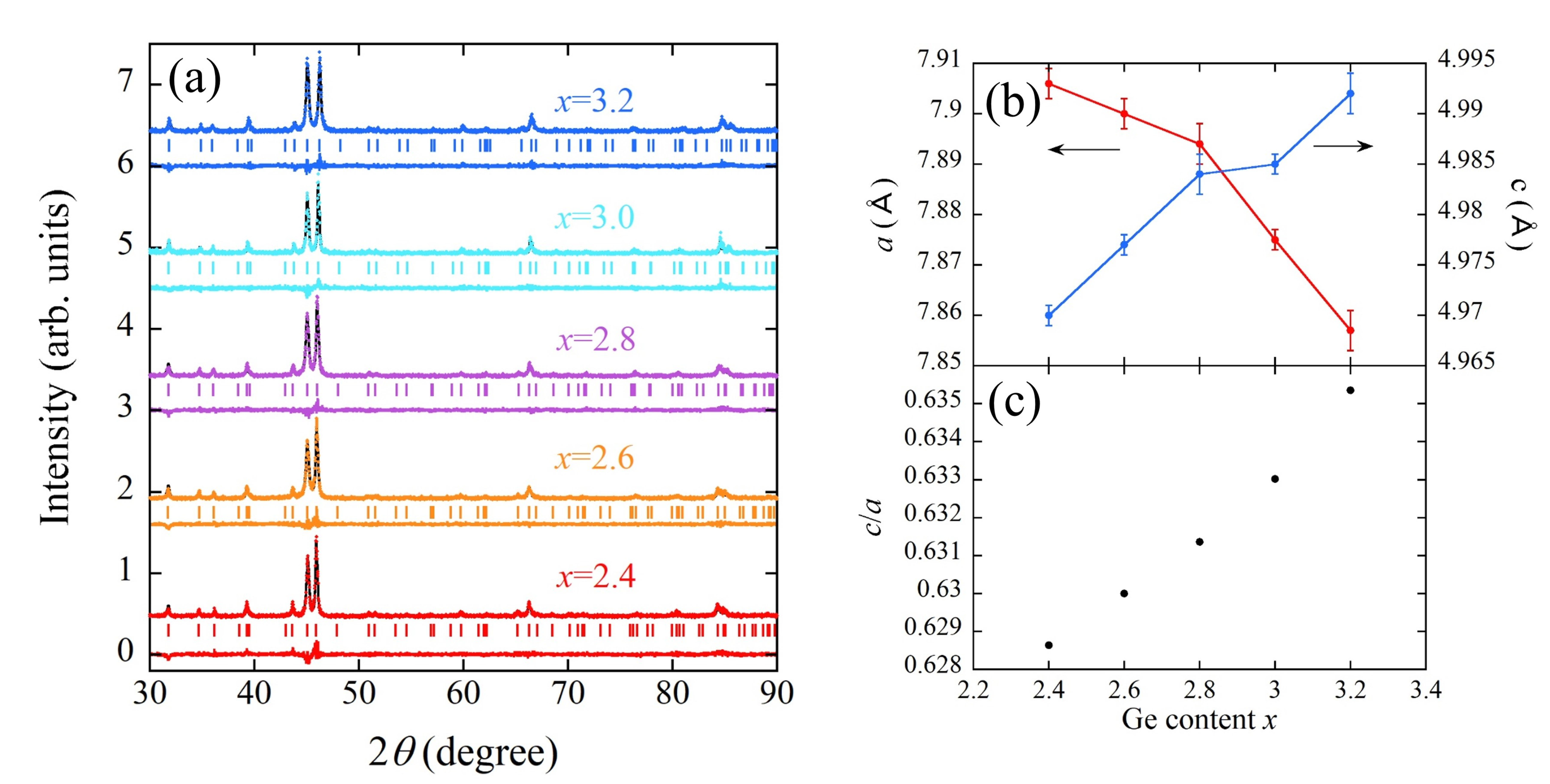}
\caption{\label{fig2} (a) XRD patterns of Co$_{6.2}$Ga$_{3.8-x}$Ge$_{x}$ ($x$=2.4 to 3.2). In each dataset, the observed ($\circ$) and calculated (solid line) XRD patterns are shown at the top. The difference between the observed and calculated XRD patterns is shown at the bottom. The tick marks indicate the positions of Bragg reflections for Co$_{6.2}$Ga$_{3.8-x}$Ge$_{x}$. (b) Ge content dependence of lattice parameters. (c) $c/a$ ratio vs. Ge content plot. }
\end{figure}

\begin{table}
\centering
\caption{\label{tab2}Lattice parameters $a$ and $c$, $T_\mathrm{C}$, $\mu_\mathrm{eff}$, $\theta_\mathrm{W}$, $H_\mathrm{c}$ at 2 K, and room temperature $\rho$ value $\rho$ (RT) of Co$_{6.2}$Ga$_{3.8-x}$Ge$_{x}$ ($x$=2.4 to 3.2).}
\footnotesize
\begin{tabular}{cccccccc}
\br
 $x$ & $a$ (\AA) & $c$ (\AA) & $T_\mathrm{C}$ (K) & $\mu_\mathrm{eff}$ ($\mu_\mathrm{B}$/Co) & $\theta_\mathrm{W}$ (K) & $H_\mathrm{c}$ at 2 K (kOe) & $\rho$ (RT) ($\mu\Omega$cm)\\
\mr
2.4 & 7.906(3) & 4.970(1) & 60.3 & 1.78 & 50 & 44 & 185   \\
2.6 & 7.900(3) & 4.977(1) & 46.4 & 1.77 & 41 & 26 & 186   \\
2.8 & 7.894(4) & 4.984(2) & 13.2 & 1.97 & -83 & 9.3 & 209   \\
3.0 & 7.875(2) & 4.985(1) & 5.8 & 1.95 & -116 & 1.4 & 126   \\
3.2 & 7.857(4) & 4.992(2) & $<$ 2 & 2.15 & -309 & 0.24 & 227   \\
\br
\end{tabular}
\end{table}
\normalsize

\begin{figure}
\centering
\includegraphics[width=0.85\linewidth]{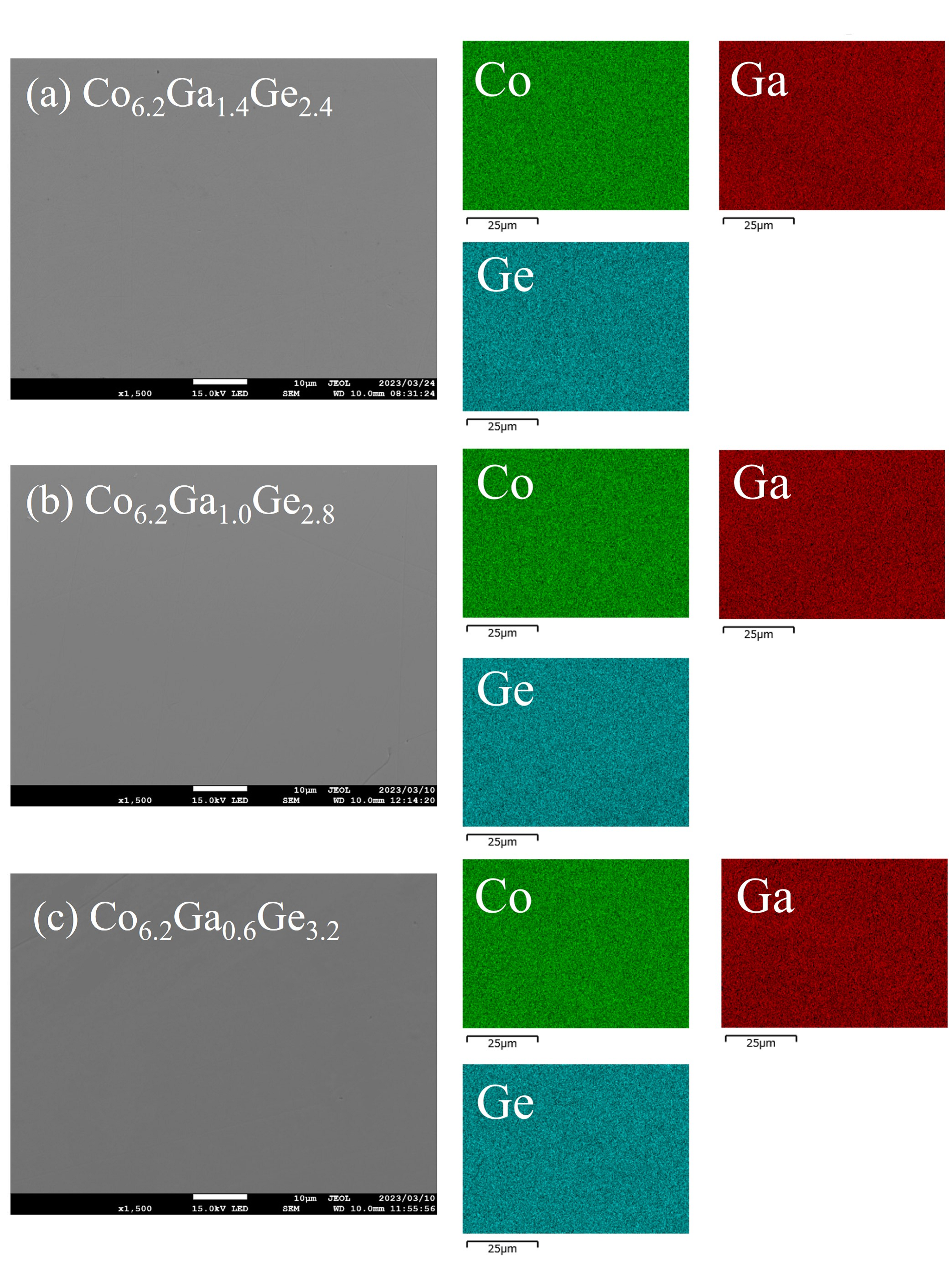}
\caption{\label{fig3} SEM images of (a) $x$=2.4, (b) $x$=2.8, and (c) $x$=3.2 for Co$_{6.2}$Ga$_{3.8-x}$Ge$_{x}$, respectively. The elemental mappings of Co, Ga, and Ge are also shown.}
\end{figure}

Figure \ref{fig2}(a) depicts the X-ray diffraction (XRD) patterns of all specimens, which are well-explained by the Fe$_{13}$Ge$_{8}$-type structure.
The hexagonal lattice parameters $a$ and $c$ are obtained with the help of a Rietveld refinement program\cite{Izumi:SSP2007,Tsubota:SR2017} and are summarized in Table \ref{tab2}.
For this purpose, we employed the RIETAN-FP program package\cite{Izumi:SSP2007}.
To ensure a full parameter fitting process of utmost precision, it is imperative for the peak intensity of XRD to exceed 10000 counts.
However, our acquired data yielded a peak intensity of at most 1000 counts.
Consequently, we adopted a strategy of fixed atomic positions, as detailed in Table \ref{tab1}, while the fitting parameters encompassed lattice parameters, background function, profile function, and scaling factor.
The lattice parameters are plotted as a function of the Ge content $x$ in Fig.\ref{fig2}(b) and reveal that the $a$-axis length reduces while the $c$-axis expands with increasing $x$. 
The $c/a$ ratio is also calculated for each sample, as displayed in Fig.\ref{fig2}(c). 
The result illustrates a systematic increase of $c/a$ as the Ge atom replaces the Ga atom gradually.
Figures \ref{fig3}(a) to (c) present scanning electron microscopy (SEM) images for several samples ($x$=2.4, 2.8, and 3.2).
Each non-contrast image indicates a homogeneous chemical composition with no conspicuous impurity phases.
The chemical compositions, ascertained through EDX analysis, are as follows: Co$_{60.7(3)}$Ga$_{14.6(5)}$Ge$_{24.8(5)}$ for $x$=2.4, Co$_{60.9(6)}$Ga$_{10.4(7)}$Ge$_{28.7(1)}$ for $x$=2.8, and Co$_{60.3(5)}$Ga$_{6.8(5)}$Ge$_{32.9(2)}$ for $x$=3.2, respectively.
Each composition is almost identical to the starting composition. 
Furthermore, elemental mappings of Co, Ga, and Ge are also presented in Figs.\ref{fig3}(a) to (c), revealing the homogeneous distribution of constituent elements.
The SEM images indicate the smoothness of the surface, which may be advantageous for various application perspectives. 
For example, the smoothness of the surface of magnetic material improves the sensitivity in tunnel magnetoresistance sensors\cite{Yamamoto:AM2021,Nakano:APL2023}. 

\begin{figure}
\centering
\includegraphics[width=1.0\linewidth]{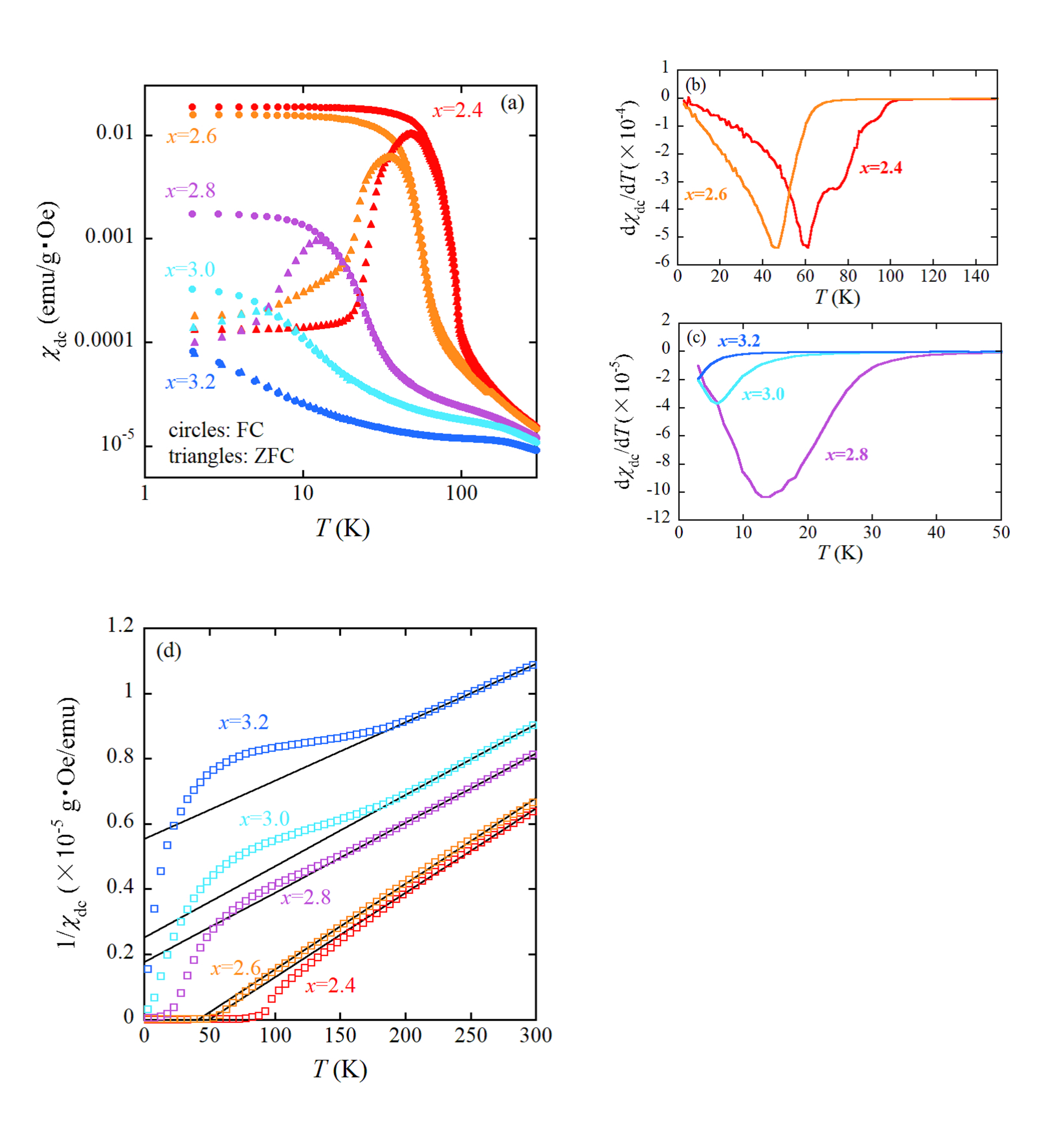}
\caption{\label{fig4} (a)Temperature dependences of $\chi_\mathrm{dc}$ of Co$_{6.2}$Ga$_{3.8-x}$Ge$_{x}$ ($x$=2.4 to 3.2) under ZFC and FC conditions. The external field is 500 Oe. Both axes are on a logarithmic scale. (b) and (c) Temperature derivative of $\chi_\mathrm{dc}$ under FC of Co$_{6.2}$Ga$_{3.8-x}$Ge$_{x}$ ($x$=2.4 to 3.2). (d) Temperature dependences of 1/$\chi_\mathrm{dc}$ of Co$_{6.2}$Ga$_{3.8-x}$Ge$_{x}$ ($x$=2.4 to 3.2). The solid lines represent the fitting results using the Curie-Weiss law.}
\end{figure}

\begin{figure}
\centering
\includegraphics[width=0.9\linewidth]{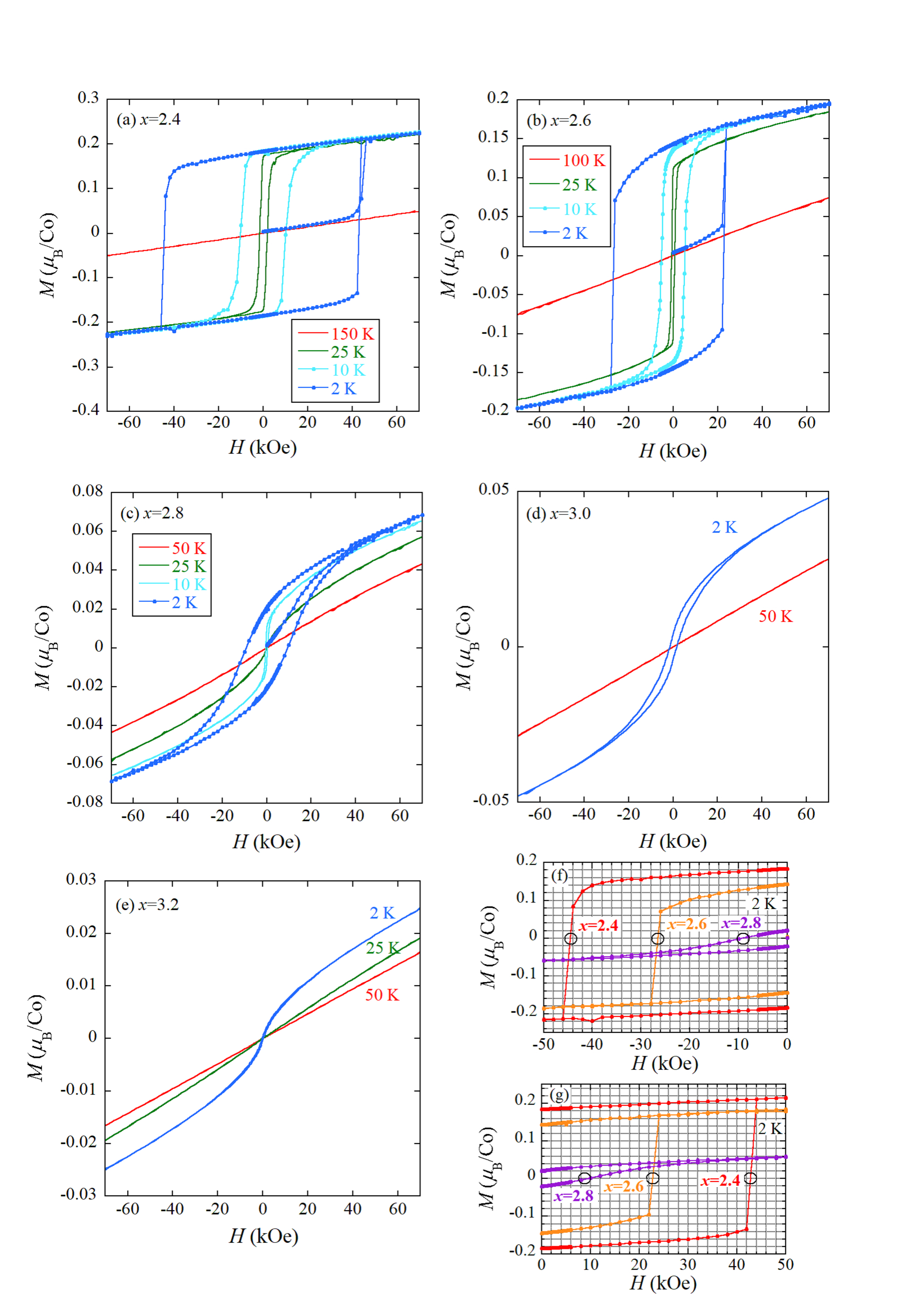}
\caption{\label{fig5} $M$-$H$ curves measured at temperatures denoted in figure for (a) $x$=2.4, (b) $x$=2.6, (c) $x$=2.8, (d) $x$=3.0, and (e) $x$=3.2 Co$_{6.2}$Ga$_{3.8-x}$Ge$_{x}$ samples. $M$-$H$ loops at 2 K around (f) negative $H_\mathrm{c}$ and (g) positive $H_\mathrm{c}$ for $x$=2.4, 2.6, and 2.8 samples.}
\end{figure}

We conducted measurements of $\chi_\mathrm{dc}$ ($T$) under both zero-field-cooled (ZFC) and field-cooled (FC) conditions, employing an external magnetic field of 500 Oe.
The temperature-dependent dc magnetic susceptibility $\chi_\mathrm{dc}$ of each sample exhibits an enhancement of $\chi_\mathrm{dc}$ in the low-temperature range, which suggests a ferromagnetic ordering, as shown in Fig.\ref{fig4}(a).
Except for the $x$=3.2 sample with magnetic ordering temperature below 2 K, the observed irreversibility between the ZFC and FC datasets is a characteristic of ferromagnetic behavior.
This phenomenon can be understood by recognizing the significant magnetic domain pinning during ZFC conditions.
The $\chi_\mathrm{dc}$ value at 2 K is heavily reduced by increasing $x$, and the $T_\mathrm{C}$ systematically decreases. 
The temperature at which the minimum temperature derivative of $\chi_\mathrm{dc}$ under FC occurs defines $T_\mathrm{C}$, shown in Table \ref{tab2} for each sample and represented in Figs.\ref{fig4}(b) and (c). 
The $x$=3.2 sample would possess $T_\mathrm{C}$ below 2 K. 
The definition of $T_\mathrm{C}$ employed in this study aligns with the methodology commonly applied in numerous ferromagnetic investigations\cite{Oikawa:APL2001,Yu:APL2003,Kitagawa:JMMM2018}. 
This approach is considered reliable, as it has been validated in some ferromagnetic materials where the $T_\mathrm{C}$ obtained through this method is consistent with $T_\mathrm{C}$ values determined via other physical quantities, such as specific heat\cite{Cao:PRB2010,Pronin:PRB2012}.
The temperature dependences of inverse $\chi_\mathrm{dc}$ are demonstrated in Fig.\ref{fig4}(d). 
In each sample, 1/$\chi_\mathrm{dc}$ follows the Curie-Weiss law expressed by $\chi_\mathrm{dc}=C/(T-\theta_\mathrm{W})$ at high temperatures, as indicated by the solid line. 
The effective magnetic moment $\mu_\mathrm{eff}$ obtained from the $C$ value and the Weiss temperature $\theta_\mathrm{W}$ are presented in Table \ref{tab2}. All $\mu_\mathrm{eff}$ values are smaller than that of an isolated Co$^{2+}$ or Co$^{3+}$ ion (3.87 or 4.90 $\mu_\mathrm{B}$/Co), which suggests an itinerant character of $d$-electrons. 
For the $x$=2.6 or 2.4 sample, the positive $\theta_\mathrm{W}$ is nearly identical to $T_\mathrm{C}$, indicating a ferromagnetic ordering. However, in the $x$=2.8 $\sim$ 3.2 samples, negative $\theta_\mathrm{W}$ values are present, indicating the dominance of AFM interaction. 
The magnetization curves exhibit hysteresis loops, which are characteristic of FM compound. 
Therefore, the samples with $x$$\geq$2.8 undergo ferrimagnetic ground states. 
Furthermore, it is notable that the frustration index $f$, defined as $|\theta_\mathrm{W}|$/$T_\mathrm{C}$, significantly increases from 6.3 to over 155 as $x$ increases from 2.8 to 3.2 in the ferrimagnetic state.
This index is grounded in the concept that $\theta_\mathrm{W}$ reflects the strength of magnetic interactions\cite{Ramirez:ARMS1994,Greedan:FO2010}.
Thus, an unfrustrated compound would attain a magnetically ordered state at $\theta_\mathrm{W}$, resulting in $f$=1. 
Geometrical frustration, such as triangular and Kagome lattices, often induces magnetic spin frustration, significantly suppressing the ordering temperature. 
Compounds of this nature frequently exhibit $f$ values exceeding 5, often surpassing 100\cite{Ramirez:ARMS1994,Greedan:FO2010}.
The observed increase in $f$ within the Co$_{6.2}$Ga$_{3.8-x}$Ge$_{x}$ series, particularly when $x$ exceeds 2.8, strongly supports the presence of spin frustration.
This finding aligns with the geometrically frustrated Kagome and triangular lattices, as depicted in Fig.\ref{fig1}.
Hence, the notable characteristic of Co$_{6.2}$Ga$_{3.8-x}$Ge$_{x}$ is its capacity for chemical manipulation of spin frustration. 

The isothermal magnetization curves (where $M$ is the magnetization and $H$ is the external field) for all samples, spanning -70 kOe to 70 kOe, are presented in Figs.\ref{fig5} (a) through (e). 
The high field $M$ at 2 K displays a sudden drop as $x$ increases from 2.6 to 2.8, which is indicative of spin compensation in the ferrimagnetic samples with $x$=2.8 $\sim$ 3.2. 
However, even in the ferromagnetic samples with $x$=2.4 or 2.6, the highest $M$ value is relatively smaller than $\mu_\mathrm{eff}$, implying the existence of a canted ferromagnetic structure. 
Notably, the $x$=2.4 and 2.6 samples exhibit a giant coercivity of 44 kOe and 26 kOe, respectively, at 2 K. 
In the $x$=2.4 or 2.6 sample, $H_\mathrm{c}$ increases significantly as the temperature drops below $T_\mathrm{C}$, and the initial magnetization curve at 2 K, exhibiting a small slope, undergoes an abrupt jump at approximately $H_\mathrm{c}$. 
This magnetization process is a typical characteristic of domain wall pinning\cite{Coey:book}. 
The ferrimagnetic samples also demonstrate hysteresis loops at low temperatures, and the corresponding $H_\mathrm{c}$ values at 2 K are provided in Table \ref{tab2}.

\begin{figure}
\centering
\includegraphics[width=0.9\linewidth]{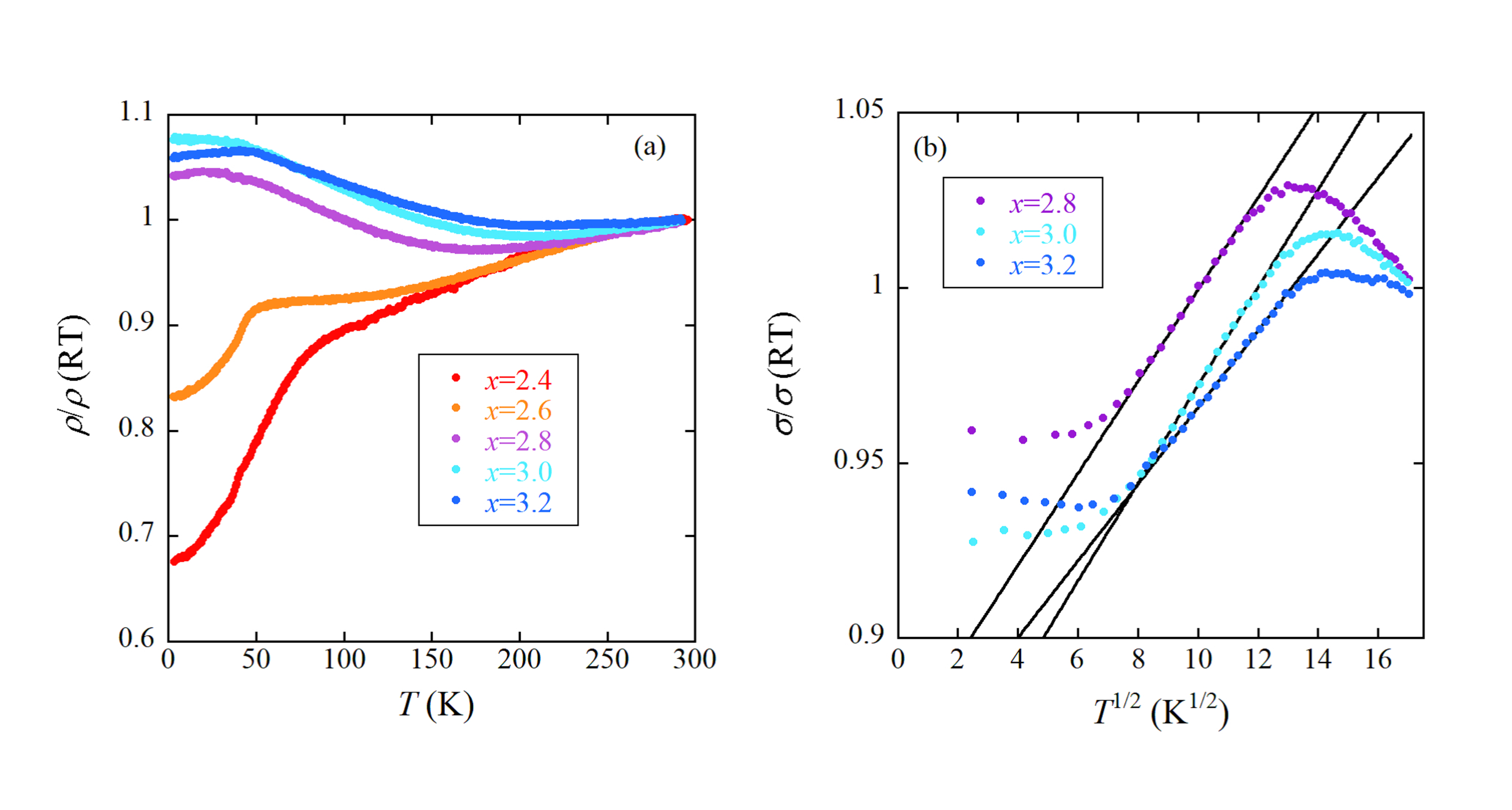}
\caption{\label{fig6} (a) Temperature dependences of $\rho$ for Co$_{6.2}$Ga$_{3.8-x}$Ge$_{x}$ ($x$=2.4 to 3.2). Each $\rho$ is normalized by the room temperature $\rho$ value $\rho$ (RT) listed in Table \ref{tab2}. (b) Temperature dependences of $\sigma$ for $x$=2.8, 3.0, and 3.2 samples plotted as a function of $T^{1/2}$. Each $\sigma$ is normalized by the room temperature $\sigma$ denoted as $\sigma$ (RT). The solid lines represent fits to the equation $\sigma_{0}+aT^{1/2}$  with the associated parameters as follows: ($\sigma_{0}$ (($\mu\Omega$cm)$^{-1}$), $a$ (($\mu\Omega$cm)$^{-1}$ K$^{-1/2}$))=(0.00416, 6.28$\times$10$^{-5}$) for $x$=2.8, (0.00663, 1.11$\times$10$^{-4}$) for $x$=3.0, and (0.00378, 4.83$\times$10$^{-5}$) for $x$=3.2, respectively.}
\end{figure}

Figure \ref{fig6}(a) displays the temperature dependences of $\rho$ in all the samples. 
The values of $\rho$ are normalized by the corresponding room temperature values listed in Table \ref{tab2}.
The metallic nature of the samples is evident from the order of magnitude of $\rho$ in each one. 
The ferrimagnetic samples exhibit a negative temperature coefficient of resistivity below around 150 K, which could be indicative of carrier localization and/or a partial opening of the gap at the Fermi surface.
For a more comprehensive understanding of carrier localization in the $x$=2.8-3.2 samples, we present the temperature-dependent electrical conductivity ($\sigma$), normalized by the room temperature $\sigma$ (designated as $\sigma$ (RT)) (=1/$\rho$ (RT)), as a function of $T^{1/2}$ in Fig.\ref{fig6}(b).
In the localization regime for each sample, $\sigma$ diminishes with decreasing temperature, adhering closely to a $T^{1/2}$ dependence.
This temperature response is explicable through a weak localization model for three-dimensional systems\cite{Lee:RMP1985}, expressed as $\sigma (T)=\sigma_{0}+aT^{1/2}$.
In this expression, the initial term represents residual conductivity, while the second term accounts for the influence of weak localization due to electron-electron interactions, with the proportional coefficient $a$. The solid lines in Fig.\ref{fig6}(b) represent fits to this model, effectively capturing the localization characteristics of Co$_{6.2}$Ga$_{3.8-x}$Ge$_{x}$ ($x$=2.8-3.2).
At lower temperatures, typically below 50 K, the experimental $\sigma$ for each sample surpasses the predicted value of the solid line. 
This observation could potentially be attributed to impurity conduction, although further investigation is warranted.

We aim to investigate the correlation between Co bond length and magnetism. 
The Bethe-Slater curve, widely used for analysing magnetism in magnetic metals, suggests that longer and shorter interatomic distances between magnetic atoms favour FM and AFM interactions, respectively\cite{Kitagawa:Metals2020-2}. 
This tendency has been observed in various intermetallic compounds\cite{Givord:1974,Kitagawa:JSSC2020}. 
The selected Co interatomic distances are listed in Table \ref{tab3}, and, taking into account the multiplicity (2 for Co1-Co1 and Co2-Co2, 6 for Co1-Co3, and 4 for Co2-Co3), Co1-Co3 and Co2-Co3 bonding likely determine the magnetic ordering type. 
The $x$-dependence of $\theta_\mathrm{W}$ in Co$_{6.2}$Ga$_{3.8-x}$Ge$_{x}$ strongly suggests the coexistence of FM and AFM interactions. 
Co1-Co3 and Co2-Co3 bonds could lead to AFM and FM interactions, respectively, although further investigation is necessary. 
When $x$ exceeds 2.8, the Co1-Co3 and Co2-Co3 bond lengths considerably decrease, consistent with the rapid predominance of AFM interaction as reflected by $\theta_\mathrm{W}$.
The coexistence of FM and AFM interactions in the Co$_{6.2}$Ga$_{3.8-x}$Ge$_{x}$ system bears similarity to artificial layered AFM-FM structures and compounds characterized by the simultaneous presence of AFM and FM phases\cite{Nogues:JMMM1999,Giri:JPCM2011}. 
In these systems, an intriguing phenomenon known as the exchange bias effect often manifests, characterized by a shift in the $M$-$H$ hysteresis loop along the magnetic field direction. 
This effect's signature can be discerned by examining the disparities between the positive and negative $H_\mathrm{c}$ values in the $M$-$H$ loop. 
Positive and negative $H_\mathrm{c}$ are the points where the $M$-$H$ loop intersects the positive and negative $x$-axis.
In Figs.\ref{fig5}(f) and (g), we have depicted the positions of the negative and positive $H_\mathrm{c}$ for the $x$=2.4, 2.6, and 2.8 samples, which exhibit larger $H_\mathrm{c}$ values at 2 K, using open circles to denote these positions. 
For the $x$=2.4 and 2.6 samples, showcasing FM ordering with weak spin frustration, the positive $H_\mathrm{c}$ values (43 kOe for $x$=2.4 and 23 kOe for $x$=2.6) are marginally lower than their corresponding negative $H_\mathrm{c}$ values (44 kOe for $x$=2.4 and 26 kOe for $x$=2.6). 
This observation supports the plausible occurrence of the exchange bias effect in the $x$=2.4 and 2.6 samples, aligning with the coexistence of FM and AFM interactions, as mentioned earlier.
Conversely, for the $x$=2.8 sample, which accompanies the spin-frustration state, the positive $H_\mathrm{c}$ value of 9.5 kOe is almost identical to the negative $H_\mathrm{c}$ value of 9.3 kOe. 
As $x$ surpasses 2.8, the rapid predominance of AFM interactions is implied by the $x$ dependence of $\theta_\mathrm{W}$. 
Consequently, the emergence of spin frustration due to the overwhelming AFM interactions may attenuate the exchange bias effect.

It should be noted that the layered hexagonal perovskite Sr$_{5}$Ru$_{4.1}$O$_{15}$, known for its giant coercive properties\cite{Ymamamoto:CM2010}, shares the same space group $P$6$_{3}$/$mmc$ (No. 194) as Co$_{6.2}$Ga$_{3.8-x}$Ge$_{x}$. 
This metallic compound exhibits a highly anisotropic crystal structure with $c/a$=4.106, wherein Ru acts as the magnetic atom and partially forms a triangular lattice. 
The analysis through Curie-Weiss fitting suggests the presence of itinerant $d$-electrons of Ru. 
The saturation moment of 0.05 $\mu_\mathrm{B}$/Ru is much smaller than the $\mu_\mathrm{eff}$ value of the Ru ion, implying weak ferromagnetism. 
The giant coercivity of Sr$_{5}$Ru$_{4.1}$O$_{15}$ arises from the large magnetocrystalline anisotropy with the geometrical frustration\cite{Ymamamoto:CM2010}.
Co$_{6.2}$Ga$_{1.4}$Ge$_{2.4}$ and Co$_{6.2}$Ga$_{1.2}$Ge$_{2.6}$, both of which are also metallic, have a $c/a$ ratio much smaller than 1.0 (as seen in Fig.\ref{fig2}(c)), indicating an anisotropic crystal structure. 
The Co atoms form Kagome and triangular lattices, and the Co magnetic moment is itinerant.
Co$_{6.2}$Ga$_{1.4}$Ge$_{2.4}$ and Co$_{6.2}$Ga$_{1.2}$Ge$_{2.6}$ exhibit relatively low saturation moments, likely originating from a canted ferromagnetic structure. 
Thus, the overall behaviour of Co$_{6.2}$Ga$_{1.4}$Ge$_{2.4}$ and Co$_{6.2}$Ga$_{1.2}$Ge$_{2.6}$ is quite similar to the magnetic and transport properties of Sr$_{5}$Ru$_{4.1}$O$_{15}$.
Therefore, we speculate that the giant coercivity in rare-earth-free magnets with the space group $P$6$_{3}$/$mmc$ can be attributed to the itinerant $d$-electrons in geometrically frustrated metals and canted spin structure, in addition to the anisotropic crystal structure.

\begin{table}
\centering
\caption{\label{tab3}Selected Co interatomic distances in Co$_{6.2}$Ga$_{3.8-x}$Ge$_{x}$ ($x$=2.4 to 3.2). The unit is \AA. \hspace{1mm} The multiplicities are 2 for Co1-Co1, 6 for Co1-Co3, 2 for Co2-Co2, and 4 for Co2-Co3, respectively. }
\begin{tabular}{ccccc}
\br
 $x$ & Co1-Co1 & Co1-Co3 & Co2-Co2 & Co2-Co3 \\
\mr
2.4 & 2.485 & 2.536 & 2.485 & 2.632   \\
2.6 & 2.489 & 2.535 & 2.489 & 2.630   \\
2.8 & 2.492 & 2.534 & 2.492 & 2.630   \\
3.0 & 2.493 & 2.530 & 2.493 & 2.625   \\
3.2 & 2.496 & 2.526 & 2.496 & 2.621   \\
\br
\end{tabular}
\end{table}

As mentioned in the introduction, CaBaCo$_{4}$O$_{7}$ is the rare example of a rare-earth-free Co-based bulk inorganic compound exhibiting a giant $H_\mathrm{c}$\cite{Caignaert:SSC2009}. 
This cobaltite also features Co-Kagome layers despite its distorted orthorhombic crystal structure. 
While its low saturation moment ($\sim$ 0.7 $\mu_\mathrm{B}$/f.u.) is akin to that of Co$_{6.2}$Ga$_{3.8-x}$Ge$_{x}$, the unique dielectric behaviour is different from the metallic transport of Co$_{6.2}$Ga$_{3.8-x}$Ge$_{x}$ ($x$=2.4 and 2.6). 
It is thus appropriate to classify Co$_{6.2}$Ga$_{3.8-x}$Ge$_{x}$ ($x$=2.4 and 2.6) as a new category of rare-earth-free Co-based inorganic compounds that exhibit giant coercivity.

A comparison of magnetic properties between Co$_{6.2}$Ga$_{3.8-x}$Ge$_{x}$ and isostructural Fe$_{3}$Ga$_{0.35}$Ge$_{1.65}$ would be of great significance. 
Fe$_{3}$Ga$_{0.35}$Ge$_{1.65}$ exhibits FM ordering below 341 K, and even at 50 K, no noticeable hysteresis loop is detected\cite{Kitagawa:JPSJ2022}. 
At 50 K, the saturation magnetization is 80 emu/g, equivalent to approximately 1.47 $\mu_\mathrm{B}$/Fe. Thus, it is likely that the itinerant nature of $d$-electrons is weakened in Fe$_{3}$Ga$_{0.35}$Ge$_{1.65}$, and the itinerant magnetic moment is essential for the manifestation of giant coercivity in transition-metal magnets with the $P$6$_{3}$/$mmc$ space group.

\section{Summary}
This study presents the transport and magnetic properties of polycrystalline Co$_{6.2}$Ga$_{3.8-x}$Ge$_{x}$ ($x$=2.4 to 3.2) with the hexagonal Fe$_{13}$Ge$_{8}$-type structure. 
The crystal structure of these materials is characterized by the presence of Kagome and triangular lattices formed by Co atoms, as well as three distinct crystallographic sites for Co atoms. 
The former feature is likely responsible for spin frustration, while the latter is significant for the competition between AFM and FM interactions. 
All samples are found to be metallic, with Co $d$-electrons exhibiting itinerancy across all $x$ values. 
As $x$ increases, the magnetic ground state changes from FM to ferrimagnetic ordering, resulting in the emergence of spin frustration, as evidenced by the large frustration index. 
Conversely, spin frustration appears to be suppressed in FM samples, although the competition of FM and AFM interactions contributes to the canted spin structure. Co$_{6.2}$Ga$_{3.8-x}$Ge$_{x}$ represents a rare system that allows for the chemical tuning of spin frustration. We note that the chemical tuning of spin frustration has not been reported in the other giant coercive materials. Notably, Co$_{6.2}$Ga$_{1.4}$Ge$_{2.4}$ and Co$_{6.2}$Ga$_{1.2}$Ge$_{2.6}$, with FM ground states, exhibit low-temperature giant coercivities comparable to the $H_\mathrm{c}$ value (30 kOe) observed in CoPt thin film studied as potential room-temperature hard magnets. This discovery of a new class of rare-earth-free Co-based compounds exhibiting giant coercivity marks a promising step toward developing rare-earth-free permanent magnets.

\section*{Acknowledgments}
J.K. is grateful for the support provided by the Comprehensive Research Organization of Fukuoka Institute of Technology. T.N. is grateful for the support from the Advanced Instruments Center of Kyushu Sangyo University

\section*{Data availability statement}   
All data that support the findings of this study are included within the article.

\section*{Author contributions}
Jiro Kitagawa: Conceptualization, Supervision, Formal analysis, Writing - original draft, Writing - reviewing \& editing. Himawari Nomura: Investigation. Terukazu Nishizaki: Investigation, Formal analysis, Writing - reviewing \& editing.

\section*{Conflicts of interest}
The authors declare no conflict of interest.

\end{document}